\documentclass[a4paper,11pt]{article}

\usepackage{jcappub}
\usepackage{amsmath}

\usepackage{graphicx}% Include figure files
\usepackage{dcolumn}% Align table columns on decimal point
\usepackage{bm}% bold math
\usepackage{color}

\newcommand{\be}{\begin{equation}}
\newcommand{\en}{\end{equation}}
\newcommand{\bea}{\begin{eqnarray}}
\newcommand{\ena}{\end{eqnarray}}

\title{Approach to exact inflation in modified Friedmann equation}
\author{Sergio del Campo}
\emailAdd{sdelcamp@ucv.cl} \affiliation{ Instituto de F\'{\i}sica,
Pontificia Universidad Cat\'{o}lica de Valpara\'{\i}so, Casilla
4059, Valpara\'{\i}so, Chile.}
\date{\today}% It is always \today, today, but any date may be explicitly specified
\begin{abstract} {We study inflationary universe models that are
characterized by a single scalar inflaton field. The study of
these models are based on two dynamical equations; one
corresponding to the Klein-Gordon equation for the inflaton field,
and the other, to a generalized Friedmann equation. After
describing the kinematics and dynamics of the models, we determine
in some detail scalar density perturbations and relic
gravitational waves. We apply this approach to the
Friedmann-Chern-Simons and the brane-world inflationary models.}
\end{abstract}

\begin{document}

\maketitle

%%%%%%%%%%%%%%%%%%%%%%%%%%%%%%%%%%%%%%%%%%%%%%%%%%%%%%%%%%%%%%%%%%%%%%%%%%%%%%%%%%%%%%%%%%%%%%%%%%%%%%%%%%%%%%%%%%%%%%%%%%%%%
\section{Introduction}\label{intro}

The early accelerated expansion of the universe (inflation)
has been proposed as a good approach for solving most of
the cosmological "puzzles" \citep{guth81,linde90a}. This
brief accelerated expansion serves, apart of solving most of the
cosmological problems, to produce the seeds
for the large scale structure formation. In fact, the present
popularity of inflation is entirely due to its
ability to generate a spectrum of density perturbations which lead
to structure formation in the universe. In essence, the conclusion
that all the observations of microwave background anisotropy
performed so far support inflation, rests on the consistency of
the anisotropy with an almost Harrison-Zel'dovich power spectrum
predicted by most of the inflationary universe scenarios
\citep{WMAP03}.

The implementation of inflationary models rests on the
introduction of a scalar {\it inflaton} field, $\phi$. The
evolution of this field becomes governed by its scalar potential,
$V(\phi)$, via the Klein-Gordon equation. In this way, this
equation of motion, together with the Friedmann equation, form the
most simple set of field equations, which can be used to study
inflationary solutions.

The previous implementation works if we give an
explicit expression for the scalar inflaton potential, $V(\phi)$.
However, in most cases result complicated to find solutions, even in
the situation in which it is applied the so-called slow-roll
approximation, where the kinetics terms is much smaller than the
potential energy, i.e. $\dot{\phi}^2 \ll V(\phi)$, together with the
approximation $\mid \ddot{\phi} \mid \ll H \mid \dot{\phi} \mid$.

Another way to find inflationary solutions, out of the slow-roll
approximation, is giving the functional form of the Hubble
parameter in term of the inflaton field, i.e. $H(\phi)$
\cite{L91}. At first glance this approach looks quite unsuitable,
since the  function $H(\phi)$ could be chosen arbitrarily. Of
course not any Hubble function, $H(\phi)$, will yield an
appropriated inflationary solution. The same it is found when the
slow-roll approximation is taken into account. There, specific
scalar potentials are chosen in order that the model could be
implemented. For instance, if the scalar potential does not
present a minimum, then, the exit from inflation becomes a
problem, since the usual reheating process can not be carried out.

It seems that the description of inflation in term of the Hubble
parameter as a function of the scalar field, i.e. $H(\phi)$, looks
"more natural"\cite{SB90}. In this way, this approach known as the
Hamilton-Jacobi formalism (H-J) provides us with a
straight-forward way of exploring the inflationary
scenario\cite{GS88,LetAl97}.

The H-J approach presents some advantages when compared with the
slow-roll approximation: first, in the H-J formalism the form of
the potential is deduced, and second, since an exact solution is
obtained, then, application to the final period of inflation is
possible, where the kinetic term of the inflaton field in the
Friedmann equation becomes important, i.e. when studying the final
stage of inflation. In this approach we can use the scalar field,
$\phi$, as a "time variable", and for that, we demand that this
field increases monotonically, i.e. its time derivative,
$\dot{\phi}$, should not change of sign along the inflationary
evolution.

In this article we would like to study the consequences that
result when considering a modified Friedmann equation expressed by
\be {\cal{F}}(H) \equiv \left(\frac{8 \pi}{3
m_{Pl}^2}\right)\rho_{_\phi} = \left(\frac{8 \pi}{3
m_{Pl}^2}\right)\left[\frac{1}{2}\dot{\phi}^2 + V(\phi)\right]
\label{2} \en
where ${\cal{F}} \geq 0$ is an arbitrary function of the Hubble
parameter $H = \dot{a}/a$, with $a$ the scale factor, the prime
represents a derivative with respect to the scalar inflaton field
$\phi$, i.e. $' \equiv \frac{d}{d\phi}$ and the dots represent
derivatives with respect to the cosmological time, $t$. Here, the
inflaton potential is expressed by $V(\phi)$ and $m_{Pl}^2 \equiv
1/G$ represents the Planck mass.

The motivation for using this kind of equation lies in the fact
that in the literature several models have been studied which can
be reduced to such modified Friedman equation.
%Some examples are the deformed Ho\v{r}ara-Lifshitz gravity, in
%which the function ${\cal{F}}(H)$ becomes, in units $c=\hbar=1$,
%$\displaystyle {\cal{F}}(H) = H^2 + \frac{H^4}{4
%\omega\,m^2_{_{Pl}}}\left[3 - 2\ln\left(\frac{4
%\pi\,m^2_{_{Pl}}}{H^2}\right)\right]$, where $\omega$ is the
%parameter of Ho\v{r}ara-Lifshitz gravity\cite{WLW11}.
In a $L(R)$-theory of gravity in which $L(R)=R-\frac{\alpha^2}{3
R}$, where $R$ is the scalar curvature and $\alpha$ is a constant
with dimension of mass square, the Friedmann equation becomes
modified by the expression
 $\displaystyle {\cal{F}}(H) =\frac{6 H^2
-\frac{\alpha}{2}}{\frac{11}{8} - \frac{9}{4
\alpha}H^2}$\cite{CDTT04}.

Also, it is possible to consider $\displaystyle {\cal{F}}(H) = H^2
-\alpha H^4$, where $\alpha$ is a constant with dimension of
$[mass]^{-2}$. There exist various forms in arriving to this
expression for ${\cal{F}}(H)$. This has been derived by
considering a quantum corrected entropy-area relation of the type
$\displaystyle S_{\cal{A}}= m_{Pl}^2 \frac{\cal{A}}{4}
-\tilde{\alpha} \ln\left({m_{Pl}^2 \frac{\cal{A}}{4}}\right)$
\cite{CCH08}, where $\cal{A}$ is the area of the apparent horizon,
and $\tilde{\alpha}$ is a dimensionless positive constant
determined by the conformal anomaly of the fields. This conformal
anomaly is interpreted as a quantum correction to the entropy of
the apparent horizon\cite{L09a}. On the other hand, this modified
Friedmann equation could be obtained when an AdS-Schwarzschild
black-hole via holographic renormalization is considered, together
with mixed boundary conditions corresponding to the Einstein field
equations in four dimension\cite{AST09}. Also, this could be
derived in terms of spacetime thermodynamics together with a
generalized uncertainly principle of quantum gravity\cite{L09b}. A
Chern-Simons type of theory yields  to this modification
too\cite{Pato}. The resulting Friedmann equation when considering
this type of modification we will call the Friedmann-Chern-Simons
equation\cite{C12}.

On the other hand, superstring and M-Theory bring the possibility
of considering our universe as a domain wall embedded in a higher
dimensional space. In this scenario the standard model of particle
is confined to the brane, while gravitation propagate into the
bulk space–time. The effect of extra dimensions induces a change
in the Friedmann equation. Here, the function ${\cal{F}}(H)$
results to be $\displaystyle {\cal{F}}(H) = \left(\frac{8 \pi
\lambda}{3 m_{_{Pl}}^2}\right)$ $\left[\sqrt{1+ \left(\frac{3
m_{_Pl}^2 }{4 \pi \lambda} \right)H^2}-1\right]$, where $\lambda$
represents the brane tension\cite{variosBW}.

Here, in this paper, after giving a general approach to the study
of inflation based on the modified Friedmann equation, we will
describe in some detail the latter two cases specified previously,
i.e. the Friedmann-Chern-Simons and the brane-world inflationary
universe models.

%%%%%%%%%%%%%%%%%%%%%%%%%%%%%%%%%%%%%%%%%%%%%%%%%%%%%%%%%%%%%%%%%%%
%%%%%%%%%%%%%%%%%%%%%%%%%%%%%%%%%%%%%%%%%%%%%%%%%%%%%%%%%%%%%%%%%%%
%%%%%%%%%%%%%%%%%%%%%%%%%%%%%%%%%%%%%%%%%%%%%%%%%%%%%%%%%%%%%%%%%%%
%%%%%%%%%%%%%%%%%%%%%%%%%%%%%%%%%%%%%%%%%%%%%%%%%%%%%%%%%%%%%%%%%%%

\section{The exact solution approach}\label{sect2}

Our study will be based on considering Eq. (\ref{2}) together with
the scalar field equation
\be \ddot{\phi} + 3 H \dot{\phi} + V'(\phi) = 0. \label{3} \en
In obtaining this latter equation we have assumed that the matter,
specified by the inflaton scalar field, enters into the action
Lagrangian in such a way that its variation in a
Friedmann-Robertson-Walker background metric leads to the
Klein-Gordon equation, expressed by Eq. (\ref{3}). Therefore, we are considering
constrained sort of models, in which the background (together with the perturbed
equation, see Eq. (\ref{27})) are not modified. In this context theory of gravity , such that
Ho\v{r}ava-Lifshitz\cite{M10} lies outside of the approach followed here.
Also, in this study we will use the scalar field $\phi$ as a "time variable".
The requirement imposed in this approach is that the scalar field
increases monotonically and its time derivative, $\dot{\phi}$,
should not change of sign along the path evolution.

From the field Eqs. (\ref{2}) and (\ref{3}) it is found that
\be \dot{\phi} = -\,\left(\frac{m_{Pl}^2}{8 \pi}\right)
{\cal{F}}_{_{,\small{H}}}\,\left(\frac{H'}{H}\right), \label{4}\en
where ${\cal{F}}_{_{,H}} \equiv d{\cal{F}}/d H$. This latter
equation allows us to write down an explicit expression  for the
scalar potential
\be V(\phi) = \frac{3 m_{Pl}^2}{8
\pi}\,{\cal{F}}\left[1-\frac{m_{Pl}^2}{48 \pi}
\left(\frac{H'\,{\cal{F}}_{_{,H}}}{H\,
\sqrt{{\cal{F}}}}\right)^2\right].\label{V1}\en

It is not hard to show that
\be a\,H = -\left(\frac{m_{Pl}^2}{8\,\pi}\right)\,
\frac{{\cal{F}}_{_{,H}}}{H}\,a'\,H',\label{6}\en
from which we get \be a(\phi) = a_i\,
\exp\left\{-\frac{8\,\pi}{m_{Pl}^2}\,
\int_{\phi_{_i}}^{\phi}{\frac{H^2}{H'\,{\cal{F}}_{_{,H}}}\,
d\phi}\right\}.\label{6}\en
where $a_i = a(\phi_{_i})$.

On the other hand, we see that the acceleration quation for the
scale factor results to be
\be \frac{\ddot{a}}{a}=
H^2\left[1-\epsilon_{_{H}}\right],\label{dda} \en
where the function $\epsilon_{_{H}}$ corresponds to
\be \epsilon_{_{H}} \equiv - \frac{d \ln {H}}{d \ln {a}} =
\left(\frac{m_{_{Pl}}^2}{8
\,\pi}\right)\frac{{\cal{F}}_{,H}}{H}\,\left(\frac{H'}{H}\right)^2.\label{epsilon}
\en
From this latter expression we can see that this definition,
called the {\it first Hubble slow-roll parameter}, gives
information about the acceleration of the universe. During
inflation we have that $\epsilon_{_{H}} <1$, and this period ends
when $\epsilon_{_{H}}$  takes the value equal to one. In the next
section we will use this parameter for describing scalar and
tensor perturbations.

One interesting quantity in characterizing inflationary universe
models is the amount of inflation. Usually, this quantity is
defined by
\be N(t) \equiv \ln {\frac{a\left(t_{e}\right)}{a(t)}},\label{N}
\en
where $a\left(t_{e}\right)$ corresponds to the scale factor
evaluated at the end of inflation. For the modified Friedmann
Equation it becomes, in terms of the scalar field
\be N(\phi) = \int_t^{t_{e}} {H \,dt} =
\left(\frac{8\,\pi}{m_{_{Pl}}^2}\right)\int^{\phi}_{\phi_{e}}
\frac{H^2}{H'\,{\cal{F}}_{,H}}\,d\phi = \int^{\phi}_{\phi_{e}}
\frac{1}{\epsilon_{_H}}\,\frac{H'}{H}\,d\phi. \label{N2} \en
Here, $\phi_{e}$ represent the value of the scalar field at the
end of inflation. Its value is determined by imposing that
$\epsilon_{_H}\left(\phi_{e}\right) = 1$.

It seems to be more appropriated to describe the amount of
inflation in terms of the comoving Hubble length, $1/(a H)$ than
in terms of the scale factor only. In this case the amount of
inflation becomes defined as\cite{LPB94}
\be \overline{\text{N}} \equiv \ln
{\frac{a\left(t_{e}\right)\,H\left(t_{e}\right)}{a(t)\,H(t)}},
\label{N3} \en
which results into
\be \overline{\text{N}}(\phi) = \int^{\phi}_{\phi_{e}}
\left(\frac{1}{\epsilon_{_{H}}}-1\right)\,\frac{H'}{H}\,d\phi.\label{N4}
\en
Note that, in general, $\overline{\text{N}}(\phi)$ is smaller that
$N(\phi)$ and only in the slow-roll limit they coincide.

In the description of inflation it is convenient to show that
their solutions are independent from their initial conditions.
This ensure the true predictive power that presents any
inflationary universe model, otherwise the corresponding physical
quantities associated with the inflationary phase, such that the
scalar or tensor spectra, would depend on these initial
conditions. Thus, with the purpose of being predictive, any
inflationary model needs that their solutions present an attractor
behavior, in the sense that solutions with different initial
conditions should tend to a unique solution\cite{SB90}.

In order to study the corresponding inflationary attractor
solutions for our case, we follow Ref. \cite{SB90}. We start by
considering a linear perturbation, $\delta H(\phi)$, around a
given inflationary solution, expressed by $H_0(\phi)$. In the
following we will refer to this quantity as the background
solution, and any quantity with the subscript zero is assumed to
be evaluated taking into account the background solution.
Therefore, at first order on $\delta H(\phi)$, we get from the
field Equations (\ref{2}) and (\ref{3}) that
\be \left.\left[1+\frac{1}{3}\epsilon_{_{H}}\left(1-
H\frac{{\cal{F}}_{_{,HH}}}{{\cal{F}}_{_{,H}}}\right)\right]\right|_0
\delta H \simeq \left.\frac{1}{3}\left(\frac{m_{_{Pl}}^2}{8
\pi}\right){\cal{F}}_{_{,H}}\frac{H'}{H^2}\right|_0\delta
H'\label{p1},\en
This latter expression can be solved for getting
\be \delta H(\phi) = \delta H (\phi_{_i})
\exp\int_{\phi_{_i}}^\phi \left.
\left(\frac{3}{\epsilon_{_{H}}}\right)
\left[1+\frac{1}{3}\epsilon_{_{H}}\left(1-H
\frac{{\cal{F}}_{_{,HH}}}{{\cal{F}}_{_{,H}}}
\right)\right]\frac{H'}{H}\right|_0 d\phi,\label{ps} \en
where $\phi_{_i}$ corresponds to some arbitrary initial value of
$\phi$.  By considering theories in which ${\cal{F}}_{_{,H}} >
H\,{\cal{F}}_{_{,HH}} $ we find that the integrand within the
exponential term will be negative, since $d \phi$ and $H'$ have
opposite signs (assuming that $\dot{\phi}$ does not change sign
due to the perturbation $\delta H$). Thus, all the linear
perturbations tend to vanish quickly\cite{LPB94}.

%%%%%%%%%%%%%%%%%%%%%%%%%%%%%%%%%%%%%%%%%%%%%%%%%%%%%%%%%%%%%%%%%%%%%%%%
%%%%%%%%%%%%%%%%%%%%%%%%%%%%%%%%%%%%%%%%%%%%%%%%%%%%%%%%%%%%%%%%%%%%%%%
%%%%%%%%%%%%%%%%%%%%%%%%%%%%%%%%%%%%%%%%%%%%%%%%%%%%%%%%%%%%%%%%%%%%%%%
%%%%%%%%%%%%%%%%%%%%%%%%%%%%%%%%%%%%%%%%%%%%%%%%%%%%%%%%%%%%%%%%%%%%%%%%
\section{Scalar and tensor perturbations}

Inflation generates perturbations through the amplification of
quantum fluctuations, which are stretched to astrophysical scales
by the brief, but rapid inflationary expansion. The simplest
models of inflation generate two types of perturbations, density
perturbations which come from quantum fluctuations of the scalar
field\cite{L80,MCh81,H82,S82}, and relic gravitational waves which
are tensor metric fluctuations\cite{G75,S79,RSV82,FP83,AW84}. The
former experience gravitational instability and lead to structure
formation\cite{MFB92}, while the latter predicts a stochastic
background of relic gravitational waves which could influence the
cosmic microwave background anisotropy via the presence of
polarization in this\cite{GW}. The upcoming experiments such as
Planck satellite will characterize polarization anisotropy to a
higher accuracy\cite{Planck}.  It is very timely to develop the
tools which can optimally utilize the polarization information to
constrain models of the early universe. Specifically, the magnetic
modes (B-modes) are signal from cosmic inflation and suggest the
presence of gravitational waves\cite{K99}.

In order to describe these perturbations let us consider the {\em
Hamilton-Jacobi slow-roll parameters}. The  {\it first slow-roll
parameter} $\epsilon_{_H}$ was already defined in the previous
section, Eq. (\ref{epsilon}).

The {\it second slow-roll parameter}, $\eta_{_H}$, is defined as
\be \eta_{_H} \equiv -\frac{d\,\ln H'}{d\,\ln a}=
\left(\frac{m_{Pl}^2}{8\,\pi}\right)\frac{{\cal{F}}_{_{,H}}}{H}
\,\frac{H''}{H}. \label{eta} \en
We should note here that both $\epsilon_{_H} $ and $\eta_{_H} $
are exact quantities, although we call them "slow-roll
parameters". In the slow-roll limit these parameters
becomes\cite{LPB94,BP95}
\begin{eqnarray}
\epsilon_{_H} & \longrightarrow & \epsilon, \nonumber \\
\eta_{_H}  & \longrightarrow & \eta - \epsilon,
\label{25}
\end{eqnarray}
where the quantities $\epsilon$ and $\eta$ are the common {\em
slow-roll parameters} which satisfies $\epsilon \ll 1$ and $\eta
\ll 1$, in agreement with the slow-roll approximation.

The evolution equation for the Fourier modes of the scalar
perturbations (quantum mode functions) at some comovil wave number
scale $k$ is governed by\cite{M85,S86}
\be \frac{d^2 u_k}{d \tau^2} + \left(k^2 -
\frac{1}{z}\frac{d^2z}{d\tau^2}\right) u_k=0, \label{27} \en
 where $\tau$ represents the conformal time defined by
 $\tau = \int{\frac{1}{a}\,dt}$ and $u_k$ corresponds to
 the Fourier transformed of the Mukanov variable, which is defined by
 $u = z \cal{R}$, with $z= a \frac{\dot{\phi}}{H}$ and
 $\cal{R}$ defining the gauge-invariant comovil curvature perturbation.
 This latter amount remains constant outside the
 horizon, i.e. metric perturbations with wavelengths larger than
 the Hubble radius will be frozen\cite{1983}.

 During inflation it is expected that $k^2 \gg
 \frac{1}{z}\frac{d^2z}{d\tau^2}$, i.e. the physical modes are
 assumed to have a wavelength much smaller than the curvature
 scale, and thus Eq. (\ref{27}) can be solved to achieve
 \be u_k(\tau) \sim e^{-ik\tau}\left(1+\frac{{\cal
 {A}}_k}{\tau}+....\right).
\label{28}
 \en
On the other hand, when $k^2  \ll
 \frac{1}{z}\frac{d^2z}{d\tau^2}$, we have that the physical modes
 correspond to wavelengths much bigger than the curvature
 scale.

 The mass term $\frac{1}{z}\frac{d^2z}{d\tau^2}$ becomes in our case
 \begin{eqnarray}
\frac{1}{z}\frac{d^2z}{d\tau^2} &= & 2 a^2 H^2\left\{ 1
-\frac{3}{2}\eta_{_H} + \frac{1}{2}\epsilon_{_H}\left(5-3 H
\frac{{\cal{F}}_{_{,HH}}}{{\cal{F}}_{_{,H}}}\right)
    \right. \nonumber \\ && \left. + \epsilon_{_H}^2\left(1- H \frac{{\cal{F}}_{_{,HH}}}
{{\cal{F}}_{_{,H}}}+\frac{1}{2} H^2
\frac{{\cal{F}}_{_{,HHH}}}{{\cal{F}}_{_{,H}}}\right) -
\frac{1}{2}\epsilon_{_H}
 \eta_{_H}\left(3-2 H \frac{{\cal{F}}_{_{,HH}}}{{\cal{F}}_{_{,H}}}\right) +
 \frac{1}{2}\eta_{_H}^2  \right. \nonumber \\
& &\hspace{6.0cm} {\left.
 + \frac{1}{2H}\dot{\epsilon}_{_H}\left(3-2 H \frac{{\cal{F}}_{_{,HH}}}
 {{\cal{F}}_{_{,H}}}\right) - \frac{1}{2 H}\dot{\eta}_{_H}
 \right\}}.
\label{29}
 \end{eqnarray}
For ${\cal{F}} (H)=H^2$ we obtain that\cite{SL93}
 $$
\frac{1}{z}\frac{d^2z}{d\tau^2} = 2 a^2 H^2\left\{ 1 +
\epsilon_{_H}
 -\frac{3}{2}\eta_{_H}- \frac{1}{2}\epsilon_{_H}
 \eta_{_H} + \frac{1}{2}\eta_{_H}^2  +
\frac{1}{2H}\dot{\epsilon}_{_H} - \frac{1}{2 H}\dot{\eta}_{_H}
 \right\}.
$$

It has long been known that Eq. (\ref{27}) could be solved exactly
in the case in which the mass term $
\frac{1}{z}\frac{d^2z}{d\tau^2}$ is proportional to $\tau^{-2}$,
in which case this equation reduces to a Bessel equation, where
the standard solution becomes $u_k \sim \sqrt{-k \tau} H_{\nu}(-k
\tau)$, with $H_{\nu}$ the Hankel function of first kind, and the
parameter $\nu$ depends on the slow-roll parameter $\epsilon$ via
$\nu = 3/2 + \epsilon/(1-\epsilon)$. For instance, this occurred
in the case of standard Friedmann equation and the scale factor,
$a(t)$, expands as a power law, i.e. $a(t) \sim t^p\,\, (p
> 1)$, in which case it is obtained that $\epsilon_{_H} =
\eta_{_H} = \text{Constant}$\cite{LS92}. Others solutions which
are far from the slow-roll approximation are described in Ref.
\cite{K97}.

Immediately we get an explicit expression for $u_k$, we can obtain the power
spectrum, which is defined in terms of the two point correlation function as
\be
{\cal{P}}_{\cal{R}}(k)=\frac{k^3}{2\pi^2}< {\cal{R}}_{{\overrightarrow{k}'}}
{\cal{R}}_{{\overrightarrow{k}} }>
\delta (\overrightarrow{k}'+\overrightarrow{k} ),%+{)which
\label{28}
\en
which in terms of the $u_k$ and $z$ it becomes
\be
 {\cal{P}}_{\cal{R}}(k)=\frac{k^3}{2\pi^2}\left|\frac{u_k}{z}\right|^{^{2}}.
\label{29} \en

In order to obtain $u_k$ by solving equation (\ref{27}), we need
to impose some boundary conditions. These asymptotic conditions
are usually taken to be the so-called Bunch-Davies vacuum
states\cite{K12}
\be
u_k \rightarrow \left\{ \begin{array}{lll}
\frac{1}{\sqrt{2k}} e^{-i k \eta} &\hspace{0.5cm} $as$& -k \eta
\longrightarrow \infty, \\
{\cal {A}}_k z &\hspace{0.5cm} $as$  & -k \eta \longrightarrow 0.
\end{array}\right.
\label{30} \en
This ensures that perturbations that are generated well inside the
horizon, i.e. in the region where $k \ll aH$, the modes approach
plane waves and those that are generated well outside the horizon,
i.e. in the region where $k \gg aH$, remain unchanged.

The description of the primordial curvature perturbation presents
a standard result given by\cite{H82,S82,varios02}
\be {\cal{P}}_{\cal{R}}(k) =\left.
\left(\frac{H}{|\dot{\phi}|}\right)^2\,\left(\frac{H}{2\,
\pi}\right)^2\right|_{aH=k}\label{35}
\en
This perturbation is in general a function of the wave number k,
which is evaluated for $aH=k$, i.e. when a given mode crosses
outside the horizon during inflation. Since the modes do not
evolve outside the horizon, the amplitude of the modes when they
cross back inside the horizon, coincides with the value that they
had when they left the horizon.

By using the primordial scalar perturbations we can introduce the
scalar spectral index $n_s$ defined by
\be n_s - 1 \equiv \frac{d \ln {\cal{P}}_{\cal{R}}}{d \ln {k}}.
\label{36}\en
This quantity becomes \be n_s - 1 = 2 \eta_{_{H}}
-2\left(3-H\,\frac{{\cal{F}}_{_{,HH}}}{{\cal{F}}_{_{,H}}}
\right)\epsilon_{_{H}}.\label{37} \en
In the same way we define the {\it running scalar spectral index},
$\alpha_s \equiv \displaystyle \frac{d n_s}{d \ln {k}} $ which
results to be given by
\be \alpha_s=2\left( 8-3 H
\frac{{\cal{F}}_{_{,HH}}}{{\cal{F}}_{_{,H}}}\right)
\epsilon_{_{H}} \eta_{_{H}} -2 \left( 9-5 H
\frac{{\cal{F}}_{_{,HH}}}{{\cal{F}}_{_{,H}}} + H^2
\frac{{\cal{F}}_{_{,HHH}}}{{\cal{F}}_{_{,H}}}\right)\epsilon_{_{H}}^2
-2\, \xi_{_{H}}, \label{38} \en
where $\xi_{_{H}}$ corresponds to the {\it third slow-roll
parameter} and becomes given by
\be \xi_{_{H}} \equiv
\left(\frac{m_{_{Pl}}^2}{4\,\pi}\right)^2\left(\frac{{\cal{F}}_{_{,H}}}{H}\right)^2
\frac{H'''\,H'}{H^2}.\label{xi} \en

In addition to the scalar curvature perturbations,
transverse-traceless tensor perturbations can also be generated
from quantum fluctuations during inflation \cite{S79,MFB92}. The
tensor perturbations do not couple to matter and consequently they
are only determined by the dynamics of the background metric, so
the standard results for the evolution of tensor perturbations of
the metric remains valid. The two independent polarizations evolve
like minimally coupled massless fields with
spectrum$^1$\footnotetext[1]{We mention here that this expression
should be implemented with a factor such that $F_\alpha^2(H/\mu)$,
where $F_\alpha^{-2}(x)=\sqrt{1+x^2} - \left(\frac{1-4\alpha
\mu^2}{1+4\alpha \mu^2}\right)x^2\,\sinh^{-1}\frac{1}{x}$, when a
braneworld with a Gauss-Bonnet term is considered\cite{DLMS04}.}
\begin{eqnarray}
\label{40} {\cal {P}}_{{\cal T}}= \frac{16
\pi}{m_{_{Pl}}^2}\left.\left(\frac{H}{2\pi}\right)^2
\right|_{aH=k}.
\end{eqnarray}

Similarly to the case of scalar perturbations, we evaluate the
expression on the right hand side of Eq. (\ref{40}) when the
comoving scale $k$ leaves the horizon during inflation.
Furthermore, we can introduce the {\it gravitational wave spectral
index} $n_{_{T}}$ defined by $\displaystyle n_{_{T}} \equiv
\frac{d \ln {{\cal {P}}_{{\cal T}}}}{d \ln {k}}$, which results to
be
\be
 n_{_{T}} = - 2\,\epsilon_{_{H}}.
\label{nt} \en
Here, we can also introduce the {\it running tensor spectral
index} $\alpha_{_T}$ defined by
\be \alpha_{_T} \equiv \frac{d n_{_T}}{d \ln k} =4 \epsilon_{_{H}}
\eta_{_{H}} -2 \left( 3- H
\frac{{\cal{F}}_{_{,HH}}}{{\cal{F}}_{_{,H}}}\right)\epsilon_{_{H}}^2\label{alpha}
\en

At this point we can define the {\it tensor-to-scalar amplitude
ratio} $\displaystyle r \equiv \frac{{\cal {P}}_{{\cal
T}}}{{\cal{P}}_{\cal{R}}}$ which becomes
\be r = 2\frac{{\cal{F}}_{_{,H}}}{H} \epsilon_{_{H}}, \label{r}
\en
and combining equations (\ref{nt}) and (\ref{r}) we find that
\be n_{_{T}} = -\left(\frac{H}{{\cal{F}}_{_{,H}}}\right) r.
\label{ntr} \en
This latter expression corresponds to the {\it inflationary
consistency condition}\cite{S85,SL93}. Note that for standard
cosmology, in which ${\cal{F}}(H)=H^2$, Eq. (\ref{ntr}) reduces to
$r=-\frac{1}{2}n_{_T}$. However, this relation could be violated
in some cases\cite{HK02, EGKSh02}. Note that this relation depends
on the kind of theory that we are dealing with.

%%%%%%%%%%%%%%%%%%%%%%%%%%%%%%%%%%%%%%%%%%%%%%%%%%%%%%%%%%%%%%%%%%%%%%%
%%%%%%%%%%%%%%%%%%%%%%%%%%%%%%%%%%%%%%%%%%%%%%%%%%%%%%%%%%%%%%%%%%%%%%%
%%%%%%%%%%%%%%%%%%%%%%%%%%%%%%%%%%%%%%%%%%%%%%%%%%%%%%%%%%%%%%%%%%%%%%%
%%%%%%%%%%%%%%%%%%%%%%%%%%%%%%%%%%%%%%%%%%%%%%%%%%%%%%%%%%%%%%%%%%%%%%%
\section{The hierarchy of the slow-roll parameters and the flow equations}

There exist a different way of studying inflationary universe
models, which is subtended by a sort of hierarchy imposed on the
slow-roll parameters\cite{HT01,K02}. In fact, the set of equations
in this approach is based on derivatives with respect to the {\it
e-folding number} over the slow-roll parameters.

We have previously introduced the slow-roll parameters, such as
$\epsilon_{_{H}}$, $\eta_{_{H}}$ and $\xi_{_{H}}$, to which we
have given a sort of hierarchy calling them {{\it first}, {\it
second} and {\it third slow-roll parameters}, respectively. Each
of these parameters is characterized by their dependence on the
order of the scalar field derivative of the Hubble ratio,
$H(\phi)$, such as $\epsilon_{_{H}} \sim (H')^2$, $\eta_{_{H}}
\sim H''$ and $\xi_{_{H}} \sim H'''$, as we can see from equations
(\ref{epsilon}), (\ref{eta}) and (\ref{xi}), respectively. It is
possible extend this definition to higher derivatives of the
Hubble parameter so that we can introduce the following parameter
\be ^l\lambda_{_H} \equiv
\left(\frac{m_{_{Pl}}^2}{4\,\pi}\right)^l\left(\frac{{\cal{F}}_{_{,H}}}{H}\right)^l
\frac{\left(H'\right)^{l-1}}{H^{^{l}}}\,\frac{{d\,}^{^{l+1}}H}{d\,\phi^{^{l+1}}}\hspace{1cm}
(l\geq 1),\label{lambda}\en
where for $l=1$ we have that $^1\lambda_{_H} \equiv \eta_{_{H}}$
and $l=2$ corresponds to $^2\lambda_{_H} \equiv \xi_{_{H}}$.

It not hard to show that the following set of equations is
satisfied
 \begin{eqnarray}
 \label{007}
 \frac{d \epsilon_{_{H}}}{d N} & = & \left[\left(H\frac{{\cal{F}}_{_{,HH}}}
 {{\cal{F}}_{_{,H}}}-3\right)\epsilon_{_{H}} + 2
 \eta_{_{H}}\right]\epsilon_{_{H}}, \nonumber \\
 & &\\
\frac{d\, {^l\lambda}_{_{H}}}{d N} & = &
\left[l\left(H\frac{{\cal{F}}_{_{,HH}}}{{\cal{F}}_{_{,H}}}-2\right)\epsilon_{_{H}}
+ (l-1)\eta_{_{H}}\right]\,{^l\lambda_{_{H}}} +
{^{l+1}\lambda_{_{H}}}.\hspace{1cm} (l\geq 1)\nonumber
 \end{eqnarray}
Here, it was used the relationship $\displaystyle \frac{d}{dN}
\equiv -\frac{m_{_Pl}^2}{8 \pi}\frac{{\cal{F}}_{_{,H}}}
 {H}\left(\frac{H'}{H}\right) \frac{d}{d \phi}$.

In the standard case, i.e. when ${\cal{F}}(H)=H^2$,  the above set
of equations reduces to
\begin{eqnarray} \frac{d\epsilon_{_{H}}}{d N} & = & \epsilon_{_{H}}(\sigma + 2
\epsilon_{_{H}}), \nonumber \\
\frac{d \sigma}{d N} & = & -5 \epsilon_{_{H}}\sigma -12
\epsilon_{_{H}}^2 + 2\xi_{_{H}} , \\
\frac{d\, {^l\lambda}_{_{H}}}{d N} & = & \left[\frac{l-1}{2}
\sigma + (l-2)\epsilon_{_{H}}\right]\,{^l\lambda_{_{H}}} +
{^{l+1}\lambda_{_{H}}},\hspace{1cm} (l\geq 2)\nonumber
\label{flow02}
 \end{eqnarray}
where $\sigma \equiv 2\eta_{_{H}} - 4\epsilon_{_{H}}$\cite{K02}.

 In order to solve the infinite set of equations (\ref{007}) the
 series is truncated  by imposing a vanishing value
 to a given high enough slow-roll parameter. This corresponds to
 take that $^{M+1}\lambda_{_H}=0$, for an appropriated large number
 $M$ (for instance, in the literature has been used $M=5$\cite{K02}).
 With this truncation the set of  equations has been solved both
 numerically\cite{HT01,ChE05,K02,EK03} and
 analytically\cite{L03,S07,AE08}.

 With respect to possible solutions of these equations and their relations
 with the inflationary  paradigm, it was emphasized in Ref. \cite{L03}
 that there is no a  clear connection between them. Actually, it is not
 clear that a particular solution of the flow
 equations corresponds directly to some type of inflationary
 solution. To achieve this task we must add additional ingredients to the
 corresponding  solutions. The main ingredient that has been left out
 of this scheme has been the Friedmann equation itself.
 Thus, in solving the flow equations we could get $\epsilon_{_{H}}$ as a
 function of the scalar field $\phi$ (imposing the condition that this parameter
 will satisfy the range $0 \leqslant \epsilon_{_{H}} \leqslant
 1$), and then we could get $H(\phi)$ through the following relation
\be \int_{H_i}^{H(\phi)} \sqrt{\frac{{\cal{F}}_{_{,H}}}{H^3}}\, d
H = \sqrt{\frac{8
\pi}{m_{_{Pl}}^2}}\,\int_{\phi_{_i}}^{\phi}\sqrt{\epsilon_{_{H}}(\phi)}
\,d\phi. \label{hfi} \en
Thus, in principle, we could give the function ${\cal{F}}(H)$ so
that we could get $\epsilon_{_{H}}$ as a function of $\phi$ by
solving the set of Eqs. (\ref{007}) we could obtain the Hubble
parameter, $H(\phi)$, as a function of the scalar field through
Eq. (\ref{hfi})$^2${\footnotetext[2]{In the standard case in which
${\cal{F}}=H^2$ it is obtained that
$$ H(\phi)= H_i\,\exp\left[{\sqrt{\frac{8
\pi}{m_{_{Pl}}^2}}\,\int_{\phi_{_i}}^{\phi}\sqrt{\epsilon_{_{H}}(\phi)}
\,d\phi}\right]. $$}.
With this in hand, we can obtain an explicit expression for the
scalar potential, $V(\phi)$, given by
\be V(\phi) = \left(\frac{3 m_{_{Pl}}^2}{8
\pi}\right)\,{\cal{F}}\left[1 -
\frac{1}{6}\,H\,\left(\frac{{\cal{F}}_{_{,H}}}{{\cal{F}}}\right)\,\epsilon_{_{H}}\right].
\label{V2} \en

In short, for a given function ${\cal{F}}(H)$, we could say that
the Hubble flow formalism allows us to determinate the scalar
field potential, $V(\phi)$, associated to some inflationary
universe model. In order to realize this task we first solve the
flow equations, Eqs. (\ref{007}), from which we can obtain the
first slow-roll parameter, $\epsilon_{_{H}}$, under the condition
that this parameter must satisfy the bound $0 \leqslant
\epsilon_{_{H}} \leqslant 1$ . Then, by using Eq. (\ref{hfi}), we
get the corresponding Hubble parameter as a function of the scalar
field, $H(\phi)$, from which we could obtain all the other
quantities associated to the inflationary scenario.

%%%%%%%%%%%%%%%%%%%%%%%%%%%%%%%%%%%%%%%%%%%%%%%%%%%%%%%%%%%%%%%%%%%%
%%%%%%%%%%%%%%%%%%%%%%%%%%%%%%%%%%%%%%%%%%%%%%%%%%%%%%%%%%%%%%%%%%%%
%%%%%%%%%%%%%%%%%%%%%%%%%%%%%%%%%%%%%%%%%%%%%%%%%%%%%%%%%%%%%%%%%%%%
%%%%%%%%%%%%%%%%%%%%%%%%%%%%%%%%%%%%%%%%%%%%%%%%%%%%%%%%%%%%%%%%%%%%

\section{Two interesting cases}

%%%%%%%%%%%%%%%%%%%%%%%%%%%%%%%%%%%%%%%%%%%%%%%%%%%%%%%%%%%%%%%%%%%%
%%%%%%%%%%%%%%%%%%%%%%%%%%%%%%%%%%%%%%%%%%%%%%%%%%%%%%%%%%%%%%%%%%%%

\subsection{The Friedmann-Chern-Simons model}
As stated in the introduction we would like to consider here a
model in which the Friedmann equation modifies
to$^3${\footnotetext[3]{For more details on this case see Ref.
\cite{C12}.}}
\be {\cal{F}}(H) \equiv  H^2-\alpha H^4 = \left(\frac{8
\pi}{3\,m_{Pl}^2}\right) \rho_{_\phi} \label{f},\en
where $\alpha$ is an arbitrary constant with dimension of
$m_{Pl}^{-2}$. Here, we assume that during the inflationary
evolution the Hubble parameter, $H$, satisfies the bound $H <
1/\sqrt{\alpha}$, so that the energy density associated to the
scalar field $\phi$ is positive.

It is possible to choose for the generating function a polynomial
like $H(\phi)=H_0(1+\beta \phi+\beta_2 \phi^2+....\beta_N\phi^N)$,
where $H_0$ and the different $\beta$ are constants. This sort of
solution was used to generate suitable functions of slow-roll
parameters\cite{L03}. Here, just for simplicity, and in order to
show how the this approach work, we shall take the previous
polynomial, but, up to first order in the scalar field $\phi$,
i.e. $H(\phi)=H_0(1+\beta \phi)$, with $\beta$ an arbitrary
constant with dimension of $m_{Pl}^{-1}$. In this case, the scalar
potential becomes
\begin{eqnarray}
 V(\phi) = \left(\frac{3 m_{Pl}^2
}{8\,\pi}\right) \,H_0^2\,\overline{\phi}\,\,^2\left[1-
\alpha H_0^2 {\overline{\phi}}\,\,^2\right] & \nonumber \\
& \times \left[1-\frac{m_{Pl}^2}{12
\pi}\frac{\beta^2}{\overline{\phi}\,\,^2}\left(\frac{1-2 \alpha
H_0^2 \overline{\phi}\,\,^2}{\sqrt{1- \alpha H_0^2
\overline{\phi}\,\,^2}}\right)^2\right], \label{17}
\end{eqnarray}
where $\overline{\phi} \equiv 1+\beta \phi$.

In the slow-roll approximation, i.e. where $\dot{\phi}^2 \ll
V(\phi)$ together with $\mid\ddot{\phi}\mid \ll \mid
dV(\phi)/d\phi \mid$, it is found that the scalar potential
becomes \be V(\phi)_{s-r} \simeq \left(\frac{3 m_{Pl}^2
}{8\,\pi}\right) \,H_0^2\,\overline{\phi}\,\,^2\left[1- \alpha
H_0^2 \overline{\phi}\,\,^2\right]. \label{18} \en Figure
\ref{fig1} depicts the shape of the potential for the exact case
(thick line), expressed by Eq. (\ref{17}), together with the
approximated slow-roll case, Eq. (\ref{18}). In the same figure
the dotted line represents the exact case in which the
$\alpha$-parameter is vanished.

\begin{figure}[th]
\centering
\includegraphics[width=12cm,angle=0,clip=true]{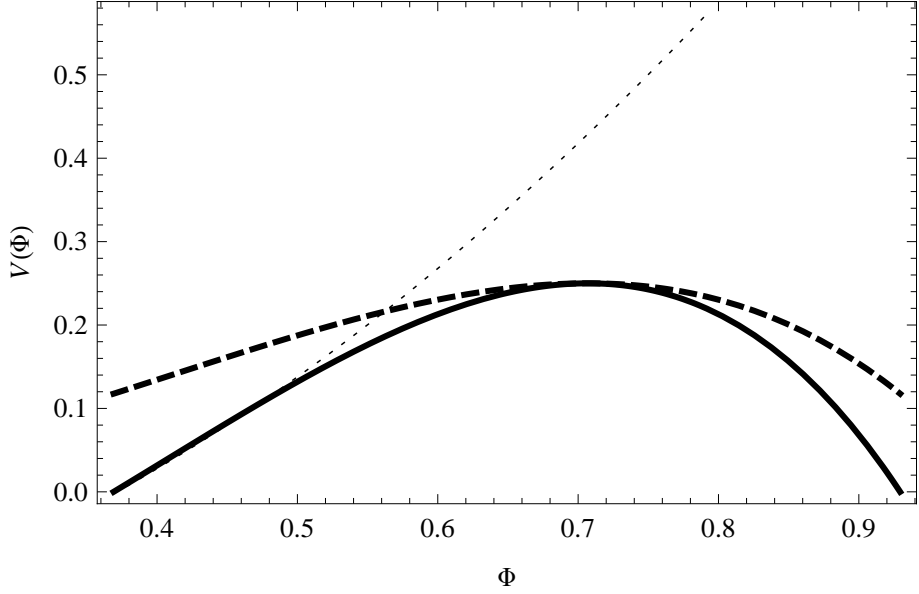}
\caption{Plots of the scalar potentials, $V(\Phi)$, as a function
of the dimensionless "scalar field", $\Phi \equiv
\sqrt{\alpha}\,H_0\,\overline{\phi}$. The thick line represents
the exact potential, expressed by Eq. (\ref{17}). Dashed line
represents the same potential, but in the slow-roll approximation,
Eq. (\ref{18}). The dotted line corresponds to the exact case, but
when $\alpha = 0$. Here we have taken $\alpha\left(\beta
H_0\right)^2 \equiv \frac{24 \pi}{9 m_{Pl}^2}$ and $V(\Phi)$ is
expressed as a multiple of the constant $ V_0\equiv \frac{3
m_{Pl}^2}{8 \pi \alpha}$.} \label{fig1}
\end{figure}

The scalar field results to be given by \be
\phi(t)=\frac{1}{\sqrt{2
\alpha}\,\beta\,H_0}\cosh\left[2\tanh^{-1}\left(\tanh\left
[\frac{1}{2}\cosh^{-1}(\sqrt{2
\alpha}\,H_0)\right]\,e^{{\cal{H}}\,(t-t_0)}\right)\right]-\frac{1}{\beta},
\label{19} \en where $\displaystyle {\cal{H}} \equiv \sqrt{2
\alpha}\,(\beta\, H_0)^2\, \frac{m_{Pl}^2}{4 \pi}$ and
$\phi(t_0)=0$. This latter expression allows us to write down the
Hubble parameter as a function of time. From this result we get
the scale factor, $a(t)$, which results to be \be a(t) = a_0
\left(\frac{\sinh\left\{2\,\tanh^{-1}\left[\tanh\left(\frac{1}{2}\cosh^{-1}(\sqrt{2
\,\alpha}\,H_0)\right)e^{{\cal{H}}(t-t_0)}\right]\right\}}{\sinh
\left[\cosh^{-1}(\sqrt{2
\,\alpha}\,H_0)\right]}\right)^{4\sqrt{2\alpha}\,{\cal{H}}}.
\label{20} \en

In order to see if this latter expression describes an accelerated
phase, for given values of the parameters, we plot in Figure
\ref{fig2} the deceleration parameters $q$, which is defined as
$q= -\frac{\ddot{a}\,a}{\dot{a}^2}$. For this plotting, we have
taken the value $\sqrt{\alpha}\,H_0 = \frac{19}{5\,\sqrt{2}}$. The
different curves correspond to different values of the exponent
that appears in the scale factor $a$, i.e.
$4\,\sqrt{2\,\alpha}\,{\cal{H}}$. These curves show that the
universe is accelerating, since the parameter $q(t)$ turns out to
be negative as time passes. Therefore, our model presents a period
of inflation, at least for the values of the parameters that we
have considered here.

\begin{figure}[th]
\centering
\includegraphics[width=12cm,angle=0,clip=true]{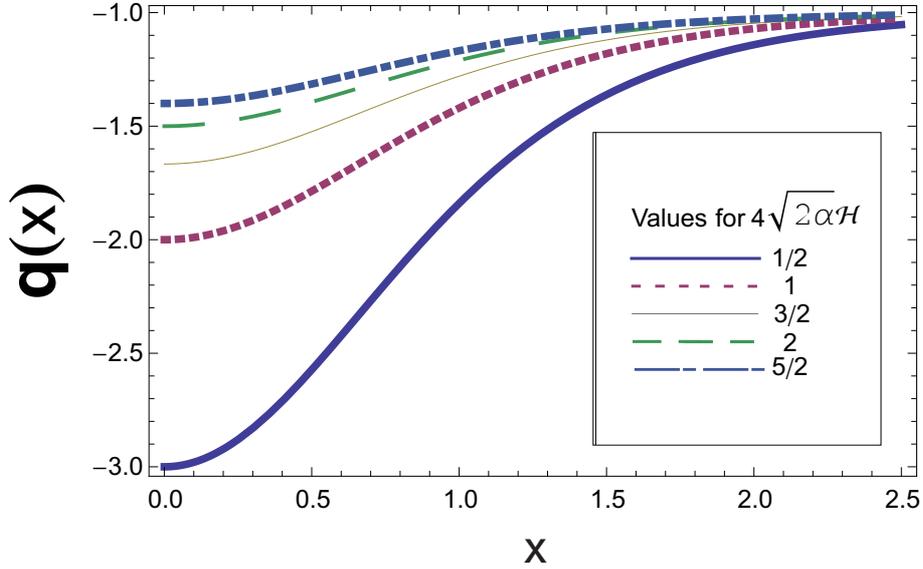}
\caption{Time evolution of the deceleration parameter, $q$, as a
function of ${\text{x}}={\cal{H}}\,(t-t_0)$. Here we have taken
the value $\sqrt{\alpha}\,H_0 = \frac{19}{5\,\sqrt{2}}$. The
thick, dotted, thin, dashed, dot-dashed lines correspond to the
values for  $4\,\sqrt{2\,\alpha}\,{\cal{H}}= 1/2; 1; 3/2; 2; 5/2$,
respectively. Note that, in all the cases the acceleration
parameters result to be negative.} \label{fig2}
\end{figure}

The amount of inflation becomes in this case
\be \overline{N}(y) = \sqrt{\gamma}\left[\frac{1}{\sqrt{y}}
+\frac{1}{\sqrt{2}}\left(\frac{1-4\gamma}{\gamma}\right)\,
\tanh^{-1}\left({\sqrt{2
y}}\right)\right\}-\overline{N}_{e},\label{N5} \en
where $y$ is a dimensionless function of the scalar field defined
by $y=\alpha H_0^2 \left(1+\beta \phi\right)^2$, $\gamma$ is a
dimensionless constant given by $\gamma \equiv
\left(\frac{m_{_{Pl}}^2}{4 \pi}\right) \alpha \left(H_0
\beta\right)^2$ and $\overline{N}_{e}$ corresponds to
$\overline{N}_{e} = \frac{1}{2}\sqrt{1+2 \gamma} + \frac{1}{2
\sqrt{2}} \left(\frac{1-4\gamma}{\sqrt{\gamma}}\right)\tanh^{-1}
\left({\sqrt{\frac{2 \gamma}{1+2 \gamma}}}\right)$.

Let us now to consider the attractor behavior of this model. By
taking into account Eq. (\ref{ps}), we get
\be \delta H(\phi) = \delta H (\phi_{_i}) \exp \left\{\frac{12
\pi}{m_{_{Pl}}^2}\int_{\phi_{_i}}^\phi
g(H_0)\frac{H_0}{H'_0}d\phi\right\}, \en
where $\phi_{_i}$ represents the initial value of the scalar field
$\phi$. The function $g(H_0)$ is given by $\left[1-2\alpha
H_0^2\left(1-\frac{2}{3}\epsilon_{_{H_0}}\alpha
H_0^2\right)\right]/\left(1-2\alpha H_0^2\right)^2$ and it is
positive for $2\alpha H_0^2 < 1$ (this makes sure that the energy
density will be positive, as we can see from Eq. (\ref{f})). Thus,
the integrand within the exponential term will be negative, due to
that $d \phi$ and $H'_0$ have contrary signs (assuming that the
perturbation $\delta H$ does not change the sign of
$\dot{\phi}$)\cite{LPB94}. In this way, all the linear
perturbations tend to vanish rapidly.

In what concern to the scalar perturbations, the scalar spectral
index parameter becomes
\be n_s - 1 = 2 \eta_{_{H}} -4\left(1-\frac{2 \,\alpha H^2}{1-2\,
\alpha H^2}\right)\epsilon_{_{H}},\label{37} \en
and the running scalar spectral index results to be
\be \alpha_s=\left( \frac{10}{1-2\,\alpha H^2}\right)
\epsilon_{_{H}} \eta_{_{H}} + \left( \frac{8}{1-2\,\alpha
H^2}\right)\epsilon_{_{H}}^2 -2\, \xi_{_{H}}^2,  \en
where  $\xi_{_{H}}$ is defined as
\be \xi_{_{H}}^2 \equiv
\left(\frac{m_{_{Pl}}^2}{4\,\pi}\right)^2\left(1-2\,\alpha
H^2\right)^2 \frac{H'''\,H'}{H^2}.\label{39} \en

Analogously, from Eq. (\ref{nt}) we find for the gravitational
wave spectral index, $n_{_{T}}$, an expression given by
\be
 n_{_{T}} = - 2\,\epsilon_{_{H}},
\label{41} \en
and we find the corresponding tensor-to-scalar amplitude ratio
\be r = 4\left(1- 2\,\alpha H^2\right) \epsilon_{_{H}}. \label{42}
\en
Bearing in mind the expression that we have required for
$H(\phi)$, we obtain a relationship between $r$ and $n_s$ given by
\be r(n_s) = \frac{(8 \gamma +1 - n_s)^2}{16 \gamma +1-n_s}, \label{rns} \en
where the dimensionless constant $\gamma$ was defines previously.
Note that we need to satisfy $n_s < 1 + 16\gamma$ in order to have
$r>0$. Thus, from this inequality we get a constraint on the
parameter $\gamma$ given by $\gamma>\frac{1}{16}|(n_s -1)|$.

Figure \ref{fig4} shows how changes $r$ as a function of $n_s$ for
two different values of the parameter $\gamma$. These values are
$\gamma =8.0 \times 10^{-3}$ and $\gamma =16.0 \times 10^{-3}$. From this figure we see that our model
can accommodate quite well the observational data. Note that this
model allows the possibility of having a Harrison-Zel'dovich
spectrum, i.e. $n_{_S}=1$,  with $r \neq 0$ as could be seen from
this plot.

\begin{figure}[th]
\centering
\includegraphics[width=12cm,angle=0,clip=true]{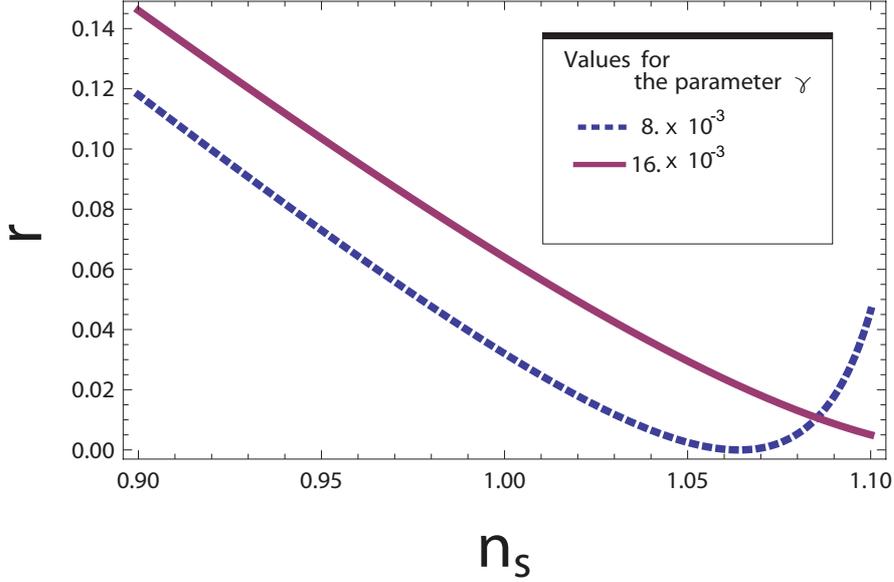}
\caption{This plot shows  the parameter $r$ as a function of the
scalar spectral index $n_s$ for two values of the constant
$\gamma=\left(\frac{m_{_{Pl}}^2}{4 \pi}\right)\alpha\left(H_0
\beta\right)^2$, as described by Eq. (\ref{rns}). Here, we have taken the values
$\gamma = 8.0 \times 10^{-3}$ and $\gamma = 16.0 \times 10^{-3}$. Note that we could
have the possibility of having a
Harrison-Zel'dovich spectrum ($n_{_S}=1$) with $r \neq 0$.}
\label{fig4}
\end{figure}

In this case the system of flow equations (\ref{007}) is reduced
to the following set of equations
 \begin{eqnarray}
 \frac{d \epsilon_{_{H}}}{d N} & = & \left[2\left(\frac{1-4\alpha H^2}
 {1-2\alpha H^2}\right)\epsilon_{_{H}} + \sigma\right]\epsilon_{_{H}}, \nonumber \\
  \frac{d \sigma}{d N} & = -&\left[6\left(\frac{2-\alpha H^2}
 {1-2\alpha H^2}\right)\epsilon_{_{H}}+\left(\frac{5-6\alpha H^2}
 {1-2\alpha H^2}\right)\sigma\right]\epsilon_{_{H}} + 2 \xi_{_H}, \label{flow03}\\
\frac{d\, {^l\lambda}_{_{H}}}{d N} & = &
\left[l\left(\frac{1-6\alpha H^2}{1-2\alpha
H^2}\right)\epsilon_{_{H}} +\frac{1}{2}
(l-1)\sigma\right]\,{^l\lambda_{_{H}}} +
{^{l+1}\lambda_{_{H}}}\hspace{1cm} (l\geq 2).\nonumber
 \end{eqnarray}
In order to solve this set of equations we need to have $H=H(N)$.
To get this, we start by considering Eq. (\ref{N}), which results
$N=N(\phi)$. Then, we need to invert this latter expression (if
possible) to obtain $\phi =\phi(N)$. Finally, with this expression
we obtain $H$ as a function of $N$, and, by introducing this
function into the flow equation we proceed to solve them.

There exist another way for getting a relationship between $H$ and
$N$. Let us assume that we really know the Hubble rate as a
function of the scale factor, i.e., we know explicitly $H(a)$.
Then, since by definition $dN=-d \ln a$, we get that $a(N)=a_e
e^{(N_e - N)}$, where $a_e$ and $N_{e}$ are the values of the
scale factor and the number of e-folding at the end of inflation.
Then, by a direct substitution of $a(N)$ on the Hubble rate $H$ it
is obtained $H(N)$.

As an example of the latter approach, let us consider the model in
which there is a smooth exit from inflation, under the so-called
{\it decaying vacuum cosmology}\cite{GMN98}. There, it was found
that the Hubble parameter as a function of the scale factor
becomes
\be H(a)= 2H_{e}\left(\frac{a_{e}^2}{a^2+a_{e}^2}\right),
\label{ha} \en
where $H_{e}=H(a_{e})$. In this case it is obtained that
\be H(N)=H_{e} \left[1-\tanh\left(N_{e}-N\right)\right].\label{hn}
\en
Let us solve numerically the set of equation (\ref{flow03}) for
the first two slow-roll parameters, $\epsilon_{_{H}}$ and
$\eta_{_H}$, when $\xi_{_H}$ is constant equal to $0.2$. Fig.
\ref{fig5} shows the numerical solutions for these two slow-roll
parameters. From this figure we see that the $\epsilon_{_{H}}$
remains almost constant (closed to zero) for a wide range of
values of $\widetilde{N}\equiv N_{e} - N$. But, for
$\widetilde{N}<1$ it increases to the value of one. Actually, for
$N=N_{e}$, i.e. at the end of inflation, $\epsilon_{_H}=1$. In the
same range, i.e. $\widetilde{N}<1$, the other slow-roll parameter,
$\eta_{_H}$, decreases from a maximum value (closed to the point
$\widetilde{N} \sim 1$) to its final value $\eta_{_H} \approx 1.2$
at the end of inflation. For this parameter, in the case
$\widetilde{N}>1$, is it observed from the figure that it
decreases lineally.

\begin{figure}[th]
\centering
\includegraphics[width=12cm,angle=0,clip=true]{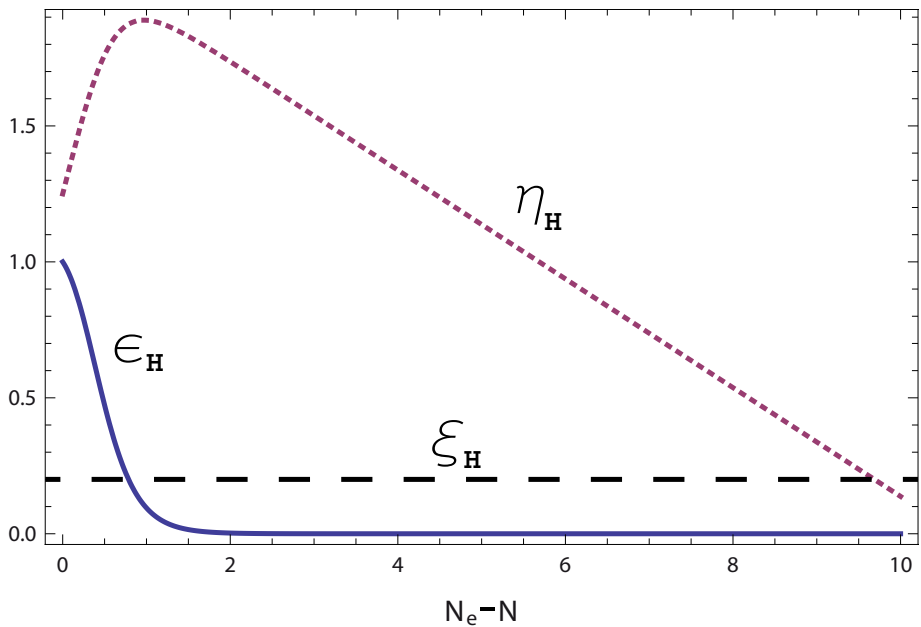}
\caption{Numerical solutions for $\epsilon_{_H}$ and $\eta_{_H}$
from the set of equations (\ref{flow03}) in the case in which
$\xi=const. =0.2$ and $H(N)=H_{e}
\left[1-\tanh\left(N_{e}-N\right)\right]$.} \label{fig5}
\end{figure}

%%%%%%%%%%%%%%%%%%%%%%%%%%%%%%%%%%%%%%%%%%%%%%%%%%%%%%%%%%%%%%%%%%%%%%%
%%%%%%%%%%%%%%%%%%%%%%%%%%%%%%%%%%%%%%%%%%%%%%%%%%%%%%%%%%%%%%%%%%%%%%%

\subsection{The brane-world model}
As was mentioned in the introduction we consider a
five-dimensional brane scenario in which the Friedmann equation is
modified to\cite{variosBW}
\be H^2=\left(\frac{8 \pi}{3
m_{_{Pl}}^2}\right)\rho_{_\phi}\left(1+\frac{\rho_{_\phi}}{2\lambda}\right),
\label{brana01} \en
where $\lambda$ represents the brane tension.

Expression (\ref{brana01}) can be written as
\be {\cal{F}}(H) \equiv  b\left[\sqrt{1+ \left(\frac{2}{b}
\right)H^2}-1\right] = \left(\frac{8 \pi}{3
m_{Pl}^2}\right)\rho_{_\phi},\label{brana02} \en
where $b$ is defined by $b \equiv \frac{8 \pi \lambda}{3
m_{Pl}^2}$.

From this latter equation and Eq. (\ref{V1}) we obtain for the
scalar potential
\be V(\phi) = \lambda \left[\sqrt{1+ \left(\frac{2}{b} \right)H^2}
- 1\right]
-\frac{2}{9}\left(\frac{\lambda}{b^2}\right)\frac{(H')^2}{[1+
\left(\frac{2}{b} \right)H^2]} \label{V2} \en

In order to obtain an explicit expression for the scalar potential
we need to introduce an explicit expression for the Hubble
parameter as a function of the scalar field. In this respect, we
borrow the expression put forward by Hawkins and Lidsey for the
Hubble parameter\cite{HL01}. Thus, we take $H(\phi) =
\sqrt{\frac{b}{2}} \left[\frac{\coth(\beta \phi)}{\sinh(\beta
\phi)}\right]$, where $\beta$ is a constant given by $\beta \equiv
\frac{\sqrt{2 \pi}\,C}{m_{_{Pl}}}$, with $C$ an arbitrary
dimensionless constant.

The scalar potential and the scale factor become
\be V(\phi) =
\frac{\lambda}{3}(6-C^2){\text{csch}}^2\left(\frac{\sqrt{2
\pi}\,C}{m_{_{Pl}}}\,\phi\right) \en
and
\be a(t) =
\frac{1}{2}bC^4\left[\left(t+\frac{4}{C^2\sqrt{b}}\right)\,t\right]^{1/C^2},
\en
respectively\cite{HL01}. Two comments are in order, first we
demand that $C$ to be less that $\sqrt{6}$ in order that the
potential is positive definite, and second, in the expression for
the scale factor it is chosen that $t_0=-\left(\frac{3
m_{_{Pl}}^2}{4 \pi \lambda C^4}\right)^{1/2}$ in order to have
$a(0)=0$. For early time it is found that $a \sim t^{1/C^2}$,
therefore for inflation to be realizable we need $C^2 < 1$.

The amount of comoving inflation becomes in this case
\be \overline{N}(x) =
\ln\left[\left(\frac{\sinh(x_{e})}{\sinh(x)}\right)^
{\frac{2}{C^2}(1-C^2)}\frac{\cosh(x_{end})}{\cosh(x)}\right]. \en
Here, $x \equiv \frac{\sqrt{2 \pi}\,C}{m_{_{Pl}}}\,\phi$ and
$x_{e} = \frac{\sqrt{2 \pi}\,C}{m_{_{Pl}}}\,\phi_{e}$ where
$\phi_{e}$ is the value of the scalar field at end of inflation,
which corresponds to $\phi_{e}=\frac{m_{_{Pl}}}{\sqrt{2
\pi}\,C}\,{\text{sech}}^{-1}\left[\frac{1}{C}\sqrt{2-C^2}\right]$.

In what concern to the attractor solutions, from Eq. (\ref{ps}) we
get that
\be \delta H(\phi) = \delta H(\phi_{_{i}})\exp
\int_{\phi_{_{i}}}^{\phi}\left.
\left[\frac{3}{\epsilon_{_H}}+\frac{2}{b}\,\frac{H^2}{1+\frac{2}{b}
\,H^2}\right]\left(\frac{H'}{H}\right)\right|_0\,d\phi.
\en
The quantity in the square bracket is positive definite, thus the
difference in sing between $H'$ and $d\phi$ makes the exponential
negative, and therefore, the exponential rapidly tends to zero,
showing the attractor feature.

Returning to the previously introduced expression for the Hubble
parameter, we obtain for the various slow-roll parameters, which
result to be given by
\be \epsilon_{_{H}}(\phi) =
\frac{C^2}{2}\,\left[1+{\text{sech}^2(\beta\,\phi)}\right],\en
\be \eta_{_{H}}(\phi) =\frac{C^2}{2}\,\left[
1+\frac{8}{3+\cosh(2\,\beta\,\phi)}\right]\en
and
\be \xi_{_H}=
\frac{C^4}{4}\,\left[1+5\,{\text{sech}}^2(\beta\,\phi)+
\frac{24}{3+\cosh(2\,\beta\,\phi)}\right].\en
By using these expressions we could obtain $n_{_s}$ and $r$ which
become
$$
n_{s}-1=
-\frac{C^2}{2}\,{\text{sech}}^2(\beta\,\phi)
\left[5+\cosh(2\,\beta\,\phi)\right]
$$
and
$$
r=2\,C^2 \tanh(\beta\,\phi),
$$
respectively.
It is not hard to show that the following relation holds
\be r=3C^2+(n_{_S}-1). \en
Now, due to observational constraint on $r$, which presents an
upper limit, $r < 0.20$ (95\% CL) from WMAP+BAO+SN\cite{WMAP07b},
where "SN" is the Constitution samples compiled in Ref.
\cite{Hetal09}, and since $n_s = 0.963 \pm 0.012$ (excluding the
Harrison-Zel'dovich spectrum in a value greater than
$3\sigma$)\cite{WMAP07a}, we get that the parameter $C$ should
satisfy the upper bound $C^2 < 0.079\pm 0.004$ in order to be in
agreement with the observational data.

On the other hand, the consistency condition in this case becomes
$r=-\frac{2}{\sqrt{1+\frac{2}{b}\,H^2}}\,n_{_T}$, where $n_{_T}$
results to be
$n_{_T}=-C^2\left[1+{\text{sech}}(\beta\,\phi)\right]$. Note that
in the limit in which $b \longrightarrow \infty$ we obtain the
standard results $r=-2\,n_{_T}$.

In what concern to the hierarchy slow-roll parameters equations we
find
 \begin{eqnarray}
 \frac{d \epsilon_{_{H}}}{d N} & = & \left[2\left(\frac{1+\frac{1}{b}\,H^2}
 {1+\frac{2}{b}\, H^2}\right)\epsilon_{_{H}} + \sigma\right]\epsilon_{_{H}}, \nonumber \\
  \frac{d \sigma}{d N} & = -&\left[\left(\frac{5+\frac{12}{b}\,H^2}
 {1+\frac{2}{b}\, H^2}\right)\sigma +12\,\epsilon_{_{H}}\right]\epsilon_{_{H}} + 2 \xi_{_H}, \label{flow04}\\
\frac{d\, {^l\lambda}_{_{H}}}{d N} & = &
\left[\left(\frac{l-2-\frac{4}{b}\,H^2}
 {1+\frac{2}{b}\, H^2}\right)\epsilon_{_{H}} +\frac{1}{2}
(l-1)\sigma\right]\,{^l\lambda_{_{H}}} +
{^{l+1}\lambda_{_{H}}}\hspace{1cm} (l\geq 2).\nonumber
 \end{eqnarray}
Following an approach analogous to the previous subsection we
solve this set numerically in the case in which the $\xi_{_H}$
parameter remains constant equal to $0.2$, and we use expression
(\ref{hn}) for the dependence of the Hubble parameter as a
function of the number of e-folding. The result is shown in Fig
\ref{fig6}. Note that $\eta_{_H}$ increases enormously close to
the end of inflation. With this parameter much greater that one
and $\epsilon_{_H}$ reaches the value equal to one at the end of
inflation, the slow-roll approximation becomes unsustainable at
the end of inflation.
\begin{figure}[th]
\centering
\includegraphics[width=12cm,angle=0,clip=true]{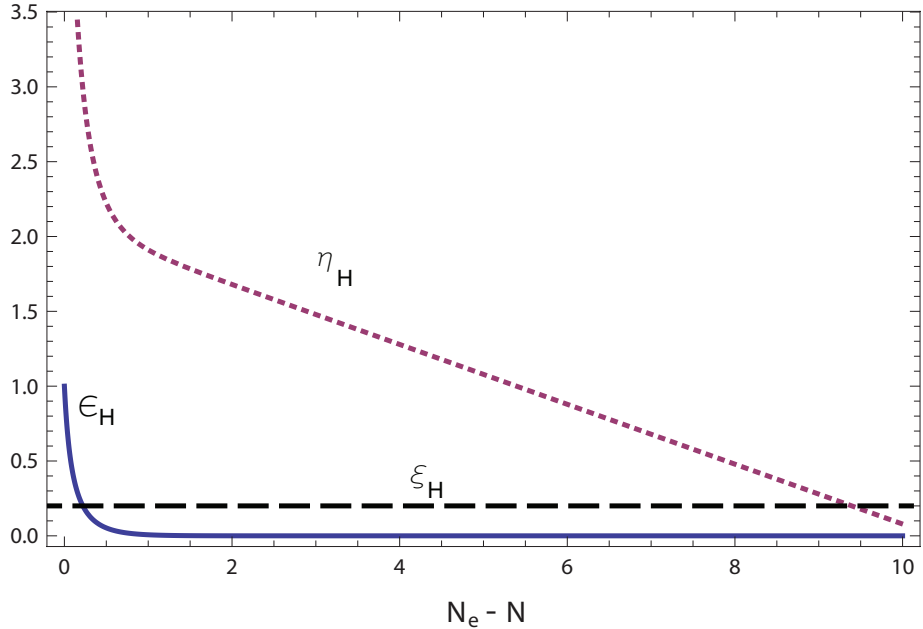}
\caption{Numerical solutions for $\epsilon_{_H}$ and $\eta_{_H}$
from the set of equations (\ref{flow04}) in the case in which
$\xi=const. =0.2$. Here, was used that $H(N)=H_e
\left[1-\tanh\left(N_e-N\right)\right]$.} \label{fig6}
\end{figure}
%

%%%%%%%%%%%%%%%%%%%%%%%%%%%%%%%%%%%%%%%%%%%%%%%%%%%%%%%%%%%%%%%%%%%%
%%%%%%%%%%%%%%%%%%%%%%%%%%%%%%%%%%%%%%%%%%%%%%%%%%%%%%%%%%%%%%%%%%%%
%%%%%%%%%%%%%%%%%%%%%%%%%%%%%%%%%%%%%%%%%%%%%%%%%%%%%%%%%%%%%%%%%%%%
%%%%%%%%%%%%%%%%%%%%%%%%%%%%%%%%%%%%%%%%%%%%%%%%%%%%%%%%%%%%%%%%%%%%

\section{Conclusion}

We have studied inflationary universe models in terms of a single
scalar field. We have applied the exact solution approach to the
modified Friedmann equations. After describing the main
characteristics of the inflationary model in general terms, we
described in some details two specific models. First, we have
studied a model characterized by a modified Friedmann equation of
the type $\displaystyle H^2 - \alpha H^4 = \left(\frac{3
m_{_{Pl}}^2}{8 \pi}\right) \rho$. Here, it was described the
kinematical evolution in the case in which the Hubble parameter
evolves as $H(\phi) = H_0 (1 + \beta \phi)$. With this at hand, we
could obtain the scalar potential, the corresponding number of
e-folding and the attractor feature of the model. For some values
of the parameters that enter into the scenario, we were able to
characterize inflationary universe models.

In what concern to the scalar and tensor perturbations we
calculated the scalar and tensor power spectrum generated by the
quantum fluctuations of the scalar and the gravitational fields.
We determined the scalar and tensor spectrum indices in term of
the so-called slow-roll parameters, $\epsilon_{_{H}}$,
$\eta_{_{H}}$ and $\xi_{_H}$. From these quantities we were able
to write down explicit expressions for the different parameters.
Moreover, the shape of the contours in the $r-n_s$ plane results
to be in agreement with those given by the WMAP 7. In fact, we
have found that the tensor-to-scalar ratio can adequately
accommodate the currently available observational data for some
values of the parameters.

In the case of the brane-world model the functional form for the
Hubble parameter was taken to be $H(\phi) = \sqrt{\frac{b}{2}}
\left[\frac{\coth(\beta \phi)}{\sinh(\beta \phi)}\right]$, where
$\beta =\frac{\sqrt{2 \pi} C}{m_{_Pl}}$. With this expression we
could determine all the kinematics and dynamics of the model. On
the other hand, the current astrophysical data put an upper bound
on the constant $C$, which becomes $C^2 < 0.079\pm 0.004$.

One of the important point that we did not considered here was the
reheating period. Since, in general terms, inflation is a period
of supercooled expansion that, when inflation ends, the
temperature of the universe needs to go up to a value such that it
coincides with that corresponding to the temperature of the
radiation epoch, and thus matching the Big Bang model. This issue,
as far as we know, has not been studied under the exact approach.
Perhaps, this study may give some insight on a deeper
understanding of the period of reheating. We hope to address this
point in the near future.

%%%%%%%%%%%%%%%%%%%%%%%%%%%%%%%%%%%%%%%%%%%%%%%%%%%%%%%%%%%%%%%%%
%%%%%%%%%%%%%%%%%%%%%%%%%%%%%%%%%%%%%%%%%%%%%%%%%%%%%%%%%%%%%%%%%
%%%%%%%%%%%%%%%%%%%%%%%%%%%%%%%%%%%%%%%%%%%%%%%%%%%%%%%%%%%%%%%%%

\begin{acknowledgments}
This work was supported by the COMISION NACIONAL DE CIENCIAS Y
TECNOLOGIA through FONDECYT Grant N$^{0}$ 1110230 and also was
partially supported by PUCV Grant N$^0$ 123.710/2011.

\end{acknowledgments}

%%%%%%%%%%%%%%%%%%%%%%%%%%%%%%%%%%%%%%%%%%%%%%%%%%%%%%%%%%%%%%%%%
%%%%%%%%%%%%%%%%%%%%%%%%%%%%%%%%%%%%%%%%%%%%%%%%%%%%%%%%%%%%%%%%%
%%%%%%%%%%%%%%%%%%%%%%%%%%%%%%%%%%%%%%%%%%%%%%%%%%%%%%%%%%%%%%%%%
%%%%%%%%%%%%%%%%%%%%%%%%%%%%%%%%%%%%%%%%%%%%%%%%%%%%%%%%%%%%%%%%%

\end{document}